\documentclass{article}
\usepackage{amssymb}
\usepackage{amsfonts}
\usepackage{amsmath}

\input{tcilatex}
\begin{document}

\title{Comment on " Solutions of the Dirac equation with an improved
expression of the Rosen-Morse potential energy model including Coulomb-like
tensor interaction "}
\author{S. Bouledjedj, A. Khodja, F. Benamira and L. Guechi \\
%EndAName
Laboratoire de Physique Th\'{e}orique, D\'{e}partement de Physique, \and %
Facult\'{e} des Sciences Exactes, Universit\'{e} des fr\`{e}res Mentouri,
\and Constantine 1, Route d'Ain El Bey, Constantine, Algeria}
\maketitle

\begin{abstract}
The Nikiforov-Uvarov polynomial method employed by Aguda to solve the Dirac
equation with an improved Rosen-Morse potential plus a Coulomb-like tensor
potential is shown inappropriate because the conditions of its application
are not fulfilled. We clarify the problem and construct the correct
solutions in the spin and pseudospin symmetric regimes via the standard
method of solving differential equations. For the bound states, we obtain
the spinor wave functions in terms of the generalized hypergeometric
functions $_{2}F_{1}(a,b,c;z)$ and in each regime we show that the energy
levels are determined by the solutions of \ a transcendental equation which
can be solved numerically.


\end{abstract}

In a recent paper \cite{Aguda} published in this journal, Aguda claimed to
have obtained the approximate analytical solutions of the Dirac equation
with an improved Rosen-Morse potential plus a Coulomb-like tensor
interaction under the condition of spin and pseudospin symmetry by using the
Nikiforov-Uvarov (NU) method. The author of this paper starts with the
following Dirac equation for a fermionic particle of mass $M$ in a mixture
of scalar and vector potentials and a Coulomb tensor potential:

\begin{equation}
\{\overrightarrow{\alpha }\text{\textperiodcentered }\overrightarrow{p}%
+\beta \lbrack M+S(r)]-i\beta \overrightarrow{\alpha }\text{%
\textperiodcentered }\widehat{r}U(r)\}\psi _{n_{r},\kappa }(\overrightarrow{r%
})=[E-V(r)]\psi _{n_{r},\kappa }(\overrightarrow{r}),  \label{a.1}
\end{equation}%
and defines the Dirac spinors as\ 
\begin{equation}
\psi _{n_{r},\kappa }(\overrightarrow{r})=\frac{1}{r}\left( 
\begin{array}{c}
F_{n_{r},\kappa }(r)Y_{jm}^{l}\left( \theta ,\phi \right) \\ 
iG_{n_{r},\kappa }(r)Y_{jm}^{\widetilde{l}}\left( \theta ,\phi \right)%
\end{array}%
\right) ,  \label{a.2}
\end{equation}%
where $F_{n,\kappa }(r)$ and $G_{n,\kappa }(r)$ are the upper and lower
component radial wave functions, respectively. Since $S(r)$, $V(r)$ and $%
U(r) $ are central potentials, it is clear that, after some simple
calculation, the components $F_{n,\kappa }(r)$ and $G_{n,\kappa }(r)$ are
determined from equation (\ref{a.1}) by the system of differential equations%
\begin{equation}
\left\{ 
\begin{array}{c}
\left( \frac{d}{dr}+\frac{\kappa }{r}-U(r)\right) F_{n_{r},\kappa
}(r)=\left( M+E_{n_{r},\kappa }-\Delta \left( r\right) \right)
G_{n_{r},\kappa }(r), \\ 
\left( \frac{d}{dr}-\frac{\kappa }{r}+U(r)\right) G_{n_{r},\kappa
}(r)=\left( M-E_{n_{r},\kappa }+\Sigma \left( r\right) \right)
F_{n_{r},\kappa }(r),%
\end{array}%
\right.  \label{a.3}
\end{equation}%
where $\Delta \left( r\right) =V(r)-S(r),$ $\Sigma \left( r\right)
=V(r)+S(r) $ and the Coulomb tensor potential $U(r)=-\frac{T}{r},$ for $%
r\geq R_{c}$ , with $T=\frac{Z_{a}Z_{b}e^{2}}{4\pi \varepsilon _{0}}$.

From these equations it follows that%
\begin{eqnarray}
&&\left[ -\frac{d^{2}}{dr^{2}}+\frac{\kappa (\kappa +1)}{r^{2}}+\left(
M+E_{n_{r},\kappa }-C_{s}\right) \Sigma \left( r\right) +\frac{T^{2}+T\left(
2\kappa +1\right) }{r^{2}}\right] F_{n_{r},\kappa }(r)  \notag \\
&=&\left[ E_{n_{r},\kappa }^{2}-M^{2}+C_{s}\left( M-E_{n_{r},\kappa }\right) %
\right] F_{n_{r},\kappa }(r)  \label{a.4}
\end{eqnarray}%
in the spin symmetry limit, i.e., when $d\Delta \left( r\right) /dr=0$ or $%
\Delta \left( r\right) =C_{s}=$constant. Here $\kappa =-l-1$ and $\kappa =l$
for $\kappa <0$ and $\kappa >0$, respectively; and $\Sigma \left( r\right) $
is the improved Rosen-Morse potential defined by

\begin{equation}
\Sigma \left( r\right) =D_{e}\left( 1-\frac{b}{e^{2r/d}+1}\right) ^{2}.
\label{a.5}
\end{equation}%
Recently, in the framework of the nonrelativistic quantum mechanics, this
potential was used to describe the vibration motion of certain diatomic
molecules in order to study their thermodynamic properties \cite%
{Song,Jia1,Jia2,Jia3}.

Similarly, from (\ref{a.3}), one has%
\begin{eqnarray}
&&\left[ -\frac{d^{2}}{dr^{2}}+\frac{\kappa (\kappa -1)}{r^{2}}-\left(
M-E_{n_{r},\kappa }+C_{ps}\right) \Delta \left( r\right) +\frac{%
T^{2}+T\left( 2\kappa -1\right) }{r^{2}}\right] G_{n_{r},\kappa }(r)  \notag
\\
&=&\left[ E_{n_{r},\kappa }^{2}-M^{2}-C_{ps}\left( M+E_{n_{r},\kappa
}\right) \right] G_{n_{r},\kappa }(r)  \label{a.6}
\end{eqnarray}%
for the pseudospin symmetry limit, i.e, when $d\Sigma \left( r\right) /dr=0$
or $\Sigma \left( r\right) =C_{ps}=$constant. In this case, $\kappa =-%
\widetilde{l}$ and $\kappa =\widetilde{l}+1$ for $\kappa <0$ and $\kappa >0$%
, respectively; and it is $\Delta \left( r\right) $ which takes the
following form of the improved Rosen-Morse potential:

{}%
\begin{equation}
\Delta \left( r\right) =D_{e}\left( 1-\frac{b}{e^{2r/d}+1}\right) ^{2}.
\label{a.7}
\end{equation}

To solve (\ref{a.4}) and (\ref{a.6}), the author of the Ref. \cite{Aguda}
made a homogenous approximation to the improved Rosen-Morse potential to
deal with the centrifugal potential term for $\kappa \neq \pm 1$ and used
the parametric generalization of polynomial NU method \cite{Nikiforov}
without considering of the conditions of its application. By defining a new
variable $s=e^{-2r/d}$ \ for $r\in \left( 0,+\infty \right) $ and $s\in
\left( 0,1\right) $ (note that the contradiction in Eq. (A10) where he says $%
s\in \left( 0,1/a_{3}\right) $ with $a_{3}=-1$ in Eq. (20)) and the
polynomial $\sigma \left( s\right) =s\left( 1-a_{3}s\right) $, Aguda asserts
that the solutions of Eqs. (18) and (25) in Ref. \cite{Aguda} can be
expressed in terms of the Jacobi polynomials (see (23) and (29) in Ref. \cite%
{Aguda}) and the relativistic energy equations are given by the Eqs. (22)
and (28) in Ref. \cite{Aguda}. However, these solutions can not be
considered as correct. On the one hand, as pointed out in Ref. \cite%
{Nikiforov} (see Eq. (17), p. 29), according to the theorem on the
orthogonality of hypergeometric-type polynomials, we note that the weight
function $\rho (s)$ does not satisfy the condition

\begin{equation}
\left. \sigma (s)\rho (s)s^{k}\right\vert _{a}^{b}=0;\text{ \ \ }%
(k=0,1,2,...).  \label{a.8}
\end{equation}%
Here $\left( a,b\right) =\left( 0,1\right) $, and the polynomial $\rho (s)$
is given by (see Appendix A in Ref. \cite{Aguda})%
\begin{equation}
\rho (s)=s^{a_{10}}(1-a_{3}s)^{a_{11}}.  \label{a.9}
\end{equation}%
This means that the solutions of equations (\ref{a.4}) and (\ref{a.6}) or
(8) and (9) in Ref. \cite{Aguda} cannot be orthogonal polynomials.

On the other hand, the solutions proposed by Aguda do not check the
condition of Hermiticity of the radial momentum operator $P_{r}=\frac{\hbar 
}{i}\frac{\partial }{r\partial r}r$ in the range $\left( R_{c},\infty
\right) $ (see Refs. \cite{Guechi1,Guechi2}). Therefore, we must discard the
solutions obtained in Ref. \cite{Aguda} entirely.

In the impossibility of applying the NU method to solve this problem, we
can, for example, directly use the standard method of solving differential
equations. In the case of the spin symmetry, by ignoring the problem of the
validity of the approximation (16) in Ref. \cite{Aguda}, we start from
equation (18) in Ref. \cite{Aguda} by changing $s$ to $(-s)$ and look for
solutions of this equation in the form

\begin{equation}
F_{n_{r},\kappa }(r)=s^{\mu }(1-s)^{\nu }f_{n_{r},\kappa }(s),  \label{a.10}
\end{equation}%
in which, on account of boundary conditions, $\mu $ has to be positive and $%
\nu $ may be a real quantity. Substituting (\ref{a.10}) into (18) in Ref. 
\cite{Aguda} and taking

\begin{equation}
\mu =\frac{d}{2}\sqrt{\beta ^{2}+\gamma +\delta D_{0}},  \label{a.11}
\end{equation}%
and 
\begin{equation}
\nu _{\pm }=\frac{1}{2}\pm \frac{1}{2}\sqrt{1+\left( \gamma b^{2}+\delta
D_{2}\right) d^{2}},  \label{a.12}
\end{equation}%
we obtain for $f_{n,\kappa }(s)$ the differential hypergeometric equations%
\begin{eqnarray}
&&\left\{ s\left( 1-s\right) \frac{d^{2}}{ds^{2}}+\left[ 2\mu +1-\left( 2\mu
+2\nu _{\pm }+1\right) s\right] \frac{d}{ds}-\left( \mu +\nu _{\pm }\right)
^{2}\right.   \notag \\
&&\left. +\left[ \beta ^{2}+\gamma b^{2}+\gamma -2\gamma b+\delta \left(
D_{0}-D_{1}+D_{2}\right) \right] \frac{d^{2}}{4}\right\} f_{n_{r},\kappa }(s)
\notag \\
&=&0,  \label{a.13a}
\end{eqnarray}%
with the following notation:%
\begin{equation}
\left\{ 
\begin{array}{c}
\beta =\sqrt{\left( M-E_{n_{r},\kappa }\right) \left( M+E_{n_{r},\kappa
}-C_{s}\right) }; \\ 
\gamma =\left( M+E_{n_{r},\kappa }-C_{s}\right) D_{e}; \\ 
\delta =\left( T+\kappa \right) \left( T+\kappa +1\right) .%
\end{array}%
\right.   \label{a.13b}
\end{equation}%
The solutions of these equations are the hypergeometric functions%
\begin{equation}
f_{n,\kappa }(s)=\mathcal{C}\text{ }_{2}F_{1}\left( \mu +\nu _{\pm }+\frac{d%
}{2}\sqrt{A},\mu +\nu _{\pm }-\frac{d}{2}\sqrt{A},2\mu +1;s\right) ,
\label{a.13c}
\end{equation}%
where 
\begin{equation}
A=\beta ^{2}+\gamma (b-1)^{2}+\delta \left( D_{0}-D_{1}+D_{2}\right) ,
\label{a.14}
\end{equation}%
and $\mathcal{C}$ is a constant factor.

Now, taking into account the formulas (see Ref.\cite{Gradshtein}, Eq.
(9.131), p. 1043),

\begin{equation}
_{2}F_{1}\left( \alpha ,\beta ,\gamma ;s\right) =\left( 1-s\right) ^{\gamma
-\alpha -\beta }\text{ }_{2}F_{1}\left( \gamma -\alpha ,\gamma -\beta
,\gamma ;s\right) ,  \label{a.15}
\end{equation}%
\begin{equation}
_{2}F_{1}\left( \alpha ,\beta ,\gamma ;s\right) =\left( 1-s\right) ^{-\alpha
}\text{ }_{2}F_{1}\left( \alpha ,\gamma -\beta ,\gamma ;\frac{s}{s-1}\right)
,  \label{a.16}
\end{equation}%
we find that the upper spinor component $F_{n_{r},\kappa }(r)$ is given by 
\begin{eqnarray}
F_{n_{r},\kappa }(r) &=&\mathcal{N}\left( e^{-2r/d}\right) ^{\mu }\left(
1+e^{-2r/d}\right) ^{-\mu -\frac{d}{2}\sqrt{A}}{}  \notag \\
&&\times \text{ }_{2}F_{1}\left( \mu +\nu _{+}+\frac{d}{2}\sqrt{A},1+\mu
-\nu _{+}+\frac{d}{2}\sqrt{A},2\mu +1;\frac{1}{e^{2r/d}+1}\right) ,  \notag
\\
&&  \label{a.17}
\end{eqnarray}%
in which $\mathcal{N}$ is a constant factor. Solution (\ref{a.17}) fulfills
the boundary condition $F_{n_{r},\kappa }(R_{c})$ $=0$, when

\begin{equation}
_{2}F_{1}\left( \mu +\nu _{+}+\frac{d}{2}\sqrt{A},1+\mu -\nu _{+}+\frac{d}{2}%
\sqrt{A},2\mu +1;\frac{1}{e^{2R_{c}/d}+1}\right) =0.  \label{a.18}
\end{equation}%
Then, the energy values for the bound states are given by the solution of
this transcendental equation (\ref{a.18}) which can be solved numerically.

In the case of pseudospin symmetry, to solve equation (25) in Ref. \cite%
{Aguda} , we proceed as before and we obtain the following expression for
the lower spinor component $G_{n_{r},\kappa }(r)$:%
\begin{eqnarray}
G_{n_{r},\kappa }(r) &=&\overline{\mathcal{N}}\left( e^{-2r/d}\right) ^{%
\overline{\mu }}\left( 1+e^{-2r/d}\right) ^{-\overline{\mu }-\frac{d}{2}%
\sqrt{A}}{}  \notag \\
&&\times \text{ }_{2}F_{1}\left( \overline{\mu }+\overline{\nu }_{+}+\frac{d%
}{2}\sqrt{A},1+\overline{\mu }-\overline{\nu }_{+}+\frac{d}{2}\sqrt{A},2%
\overline{\mu }+1;\frac{1}{e^{2r/d}+1}\right) ,  \notag \\
&&  \label{a.19}
\end{eqnarray}%
where%
\begin{equation}
\left\{ 
\begin{array}{c}
\overline{\mu }=\frac{d}{2}\sqrt{\overline{\beta }^{2}-\overline{\gamma }+%
\overline{\delta }D_{0}}, \\ 
\overline{\nu }_{+}=\frac{1}{2}\left( 1+\sqrt{1-\left( \overline{\gamma }%
b^{2}+\overline{\delta }D_{2}\right) d^{2}}\right) , \\ 
A=\overline{\beta }^{2}-\overline{\gamma }\left( b-1\right) ^{2}+\overline{%
\delta }\left( D_{0}-D_{1}+D_{2}\right)%
\end{array}%
\right.  \label{a.20}
\end{equation}%
and $\overline{\mathcal{N}}$ is a constant factor. The parameters $\overline{%
\beta },\overline{\gamma }$ and $\overline{\delta }$ involved in (\ref{a.20}%
) have in the present case the following values:

\begin{equation}
\left\{ 
\begin{array}{c}
\overline{\beta }=\sqrt{\left( M+E_{n_{r},\kappa }\right) \left(
M-E_{n_{r},\kappa }+C_{ps}\right) }, \\ 
\overline{\gamma }=D_{e}\left( M-E_{n_{r},\kappa }+C_{ps}\right) , \\ 
\overline{\delta }=\left( \kappa +T\right) \left( \kappa +T-1\right) .%
\end{array}%
\right.  \label{a.21}
\end{equation}%
Then, the energy levels of the physical system can be also found from a
numerical solution of the transcendental equation

\begin{equation}
_{2}F_{1}\left( \overline{\mu }+\overline{\nu }_{+}+\frac{d}{2}\sqrt{A},1+%
\overline{\mu }-\overline{\nu }_{+}+\frac{d}{2}\sqrt{A},2\overline{\mu }+1;%
\frac{1}{e^{2R_{c}/d}+1}\right) =0.  \label{a.22}
\end{equation}

Therefore, the analytical bound state solutions and the numerical results
given by Aguda \cite{Aguda} are not correct because \ the NU polynomial
method does not applicable to this potential type with the Dirichlet
boundary conditions. The appropriate solutions of Eqs. (\ref{a.4}) and (\ref%
{a.6}) are expressed in terms of hypergeometric series. From these, we have
shown by applying the boundary conditions that the energy levels can be
found from numerical solution of transcendental equations involving the
hypergeometric function . The numerical resolution of equations (\ref{a.18})
and (\ref{a.22}) will be developed in a future publication.

Finally, we would like to emphasize that the results of Refs. \cite%
{Song,Jia1,Jia2,Jia3} will be correct when the bound states with energy $%
E_{n_{r}}$ are determined by an transcendental equation analogous to (\ref%
{a.18}) and (\ref{a.22}). (see Ref. \cite{Khodja} for an exhaustive,
although not complete, bibliography of our works).

\end{document}